# Self-evolving ghost imaging

Baolei Liu,[1] Fan Wang,[1,2,*] Chaohao Chen,[1] Fei Dong,[3] and David McGloin[2,*]

[1]*School of Mathematical and Physical Sciences, Faculty of Science, University of Technology Sydney, Sydney, NSW 2007, Australia*
[2]*School of Electrical and Data Engineering, Faculty of Engineering and IT, University of Technology Sydney, NSW 2007, Australia*
[3]*Department of Automation, Tsinghua University, Beijing 100084, China*
*\*Corresponding author: fan.wang@uts.edu.au; david.mcgloin@uts.edu.au*

**Ghost imaging can capture 2D images with a point detector instead of an array sensor. It therefore offers a solution to the challenge of building area format sensors in wavebands where such sensors are difficult and expensive to produce and opens up new imaging modalities due to high-performance single-pixel detectors. Traditionally, ghost imaging retrieves the image of an object offline, by correlating measured light intensities and applied illuminating patterns. Here we present a feedback-based approach for online updating of the imaging result that can bypass post-processing, termed self-evolving ghost imaging (SEGI). We introduce a genetic algorithm to optimize the illumination patterns in real-time to match the object's shape according to the measured total light intensity. We theoretically and experimentally demonstrate this concept for static and dynamic imaging. This method opens new perspectives for real-time ghost imaging in applications such as remote sensing (e.g. machine vision / LiDAR systems in autonomous vehicles) and biological imaging.**

http://dx.doi.org/

Ghost imaging allows image formation with a single-pixel detector that has no spatial resolution by applying time-varying patterns to an object of interest [1, 2]. The predetermined optical patterns, generated by a scattering medium or a spatial light modulator, for example, project onto the object and a single-element photodetector records the light intensity fluctuations either transmitted or backscattered by the object. Then the image can be recovered from analysis of the correlations between the patterns and intensities. Such a single-pixel detection configuration has the potential to enable low-cost imaging in the X-ray [3], infrared [4] and terahertz wavebands [5], where detector arrays are often expensive. It also helps to enhance bioimaging deep inside scattering media, since the total fluorescence light detection scheme enables higher tolerance for signal scrambling through biological tissues, compared with conventional 2D pixel-array detection [6]. Also, it can eliminate the "out-of-focus" light through the use of sparse patterns in optical-sectioning microscopy [7]. Additionally, the time resolving ability and high-sensitivity of single-pixel detectors make ghost imaging flexible enough to straightforwardly combine with other imaging modalities [8-10]. Ghost imaging can also take advantage of the inherent sparsity of the object image and recover the image with fewer measurements than the number of its pixels using algorithms such as compressive sensing. Thus, it provides benefits such as improved frame rates with lower sampling ratios in applications such as multiphoton microscopy [11] and random-access microscopy [12] by making use of digital pattern scanning rather than mechanical raster scanning. There have also been demonstrations of ghost imaging in multimode fiber endoscopy [13] and cytometry [14].

Usually in ghost imaging more measurements lead to higher quality reconstruction but lower imaging speed and slower reconstruction time. Post-processing algorithms have been proposed to enhance the reconstruction image. Compressed sensing provides high image quality at the cost of recovery time [2] for applications that permit offline reconstruction. Other algorithms allow for differential [15], iterative [16] or Gerchberg-Saxton-like [17] data treatments, serving different functional needs. Various spatiotemporal illuminating patterns [4, 18-20], have also been proposed to either denoise the image by sampling on its inherent sparsity or enable a computationally fast algorithm. However, these methods calculate the reconstruction image after at least one sampling cycle or, in most cases, do it offline, limiting applications of ghost imaging that need continuous real-time imaging.

Here we present a feedback-based ghost imaging called self-evolving ghost imaging (SEGI), which can update the image of an object using optimized illuminating patterns generated in real-time with only of tens of measurements. By highlighting the inherent reciprocity between the total light intensity signal from an unknown object and illumination patterns, we can create evolved patterns that converge towards the image of an object as the iterated sampling loops. We introduce a genetic algorithm that has been used in wavefront optimization to focus light through scattering media [21, 22] to adaptively optimize the patterns in every generation of SEGI. Our technique enables instant imaging without the postponed computation required by other ghost imaging schemes. Further, we adapt the image denoising method, block-matching and collaborative filtering (BM3D), to numerically denoise the raw images to enhance image quality [23].

In typical ghost imaging, the illumination patterns $I_i(x,y)$ are designed prior to an experiment taking place and projected by a spatial light modulator, such as a digital micromirror device (DMD). Then the signal light intensities $S_i$ can be recorded by a bucket detector such as a photodiode, photomultiplier tube, or by pixel binning in a charge-coupled device, acting as a single-pixel detector, with:

$$S_i = \int I_i(x,y) \cdot O(x,y) \tag{1}$$

where $O(x,y)$ is the geometric function of the object. A traditional correlation strategy can retrieve the ghost image $R(x,y)$ after a sequence of measurements by

$$R(x,y) = \langle SI(x,y) \rangle - \langle S \rangle \langle I(x,y) \rangle, \tag{2}$$

where $\langle \cdots \rangle$ is the ensemble average over the distribution of patterns.

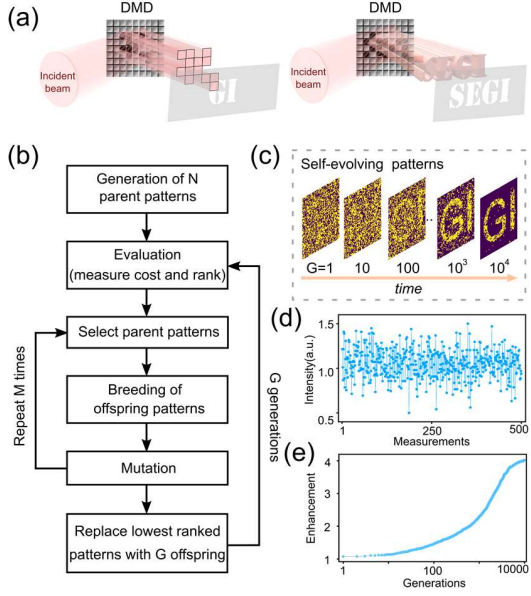

Fig. 1. Working principle of SEGI. (a) Projected patterns in ghost imaging and SEGI. (b) Block diagram showing the steps of SEGI. A population of random parent patterns are generated to illuminate the object. After the cost function measurement, these patterns are ranked for the breeding of $N$ offspring patterns. Each new pattern is created by combining the *ma* and *pa* patterns with a random breeding template, followed with a mutation operation of the pixel values. The offspring patterns replace the $S$ lowest ranks members of the parent generation. The steps are repeated in every generation. (c) Example of SEGI results from increased numbers of generation with a population of 30. G is the generation number. (d) Fluctuating intensity distribution of ghost imaging with random patterns, as a comparison to the enhanced cost function outputs (e) derived from the evolving patterns in SEGI.

In SEGI, the illuminating patterns are dynamically updated towards the shape of the target, compared with the predetermined ones in conventional ghost imaging, as shown in Fig. 1(a). A genetic algorithm, which uses principles inspired in nature to "evolve" toward a best solution, is used to iteratively optimize the patterns through operations of breeding and mutation according to measured intensities and parent patterns [21, 22], as depicted in Fig. 1(b). Initially, a population of $N$ parent patterns ($I_i^1(x,y), i = 1,2,\cdots,N$) is randomly generated for projection onto the object. Then, they are ranked according to the cost function (CF) normalized by the initial parents as defined below:

$$CF(i,g) = \frac{(S_i^g)^k \cdot \langle \sum I_i^1(x,y) \rangle}{\langle (S_i^1)^k \rangle \cdot \sum I_i^g(x,y)}, \tag{3}$$

where $\langle (S_i^1)^k \rangle$ is measured total light intensity corresponding to the initial patterns, $\langle \sum I_i^1(x,y) \rangle$. $\sum I_i^g(x,y)$ is the pattern weight determined by summing the pixel values and $S_i$ is the corresponding light intensity in the $g$ ($g \in [1,\ldots,G]$, G is the final generation number.)

generation. $k$ is a weight coefficient for $S_i^g$. Each offspring pattern is created from two random chosen parent patterns, *pa* and *ma*, from a random binary template $T$: Offspring = ma $\cdot T$ + pa $\cdot (1 - T)$, according to the cost function, with higher CF values indicating higher rankings and a correspondingly higher probability to be chosen. Similar to the natural evolution principle, the offspring pattern will mutate at partial pixels, with a decreasing mutation rate along with the generation number. $M$ mutated offspring patterns are generated by repeating the process (from parent selecting to mutation) $M$ times, here $M = N/2$. These offspring patterns will replace the $M$ lowest ranked patterns in the last generation $N$ patterns to create a new generation of $N$ patterns that self-evolve towards the object image.

Fig. 1(c) illustrates SEGI with letters 'GI' as the object in a 64x64-pixel image. With $N = 30$, a rough outline is shown by the 100th generation, and continuously improves with $G$. Compared with the fluctuating intensity distribution (Fig. 1(d)) seen in conventional ghost imaging, the enhancement of CF, during the optimization of SEGI, sustainably increases and is quadrupled by the 10000th generation with a high-quality image (Fig. 1(e)). Note that a useful generation number is not necessarily as large as 10000, especially for dynamic imaging where the structural continuity between the multiple frames can serve as a priori knowledge to assist pattern updates.

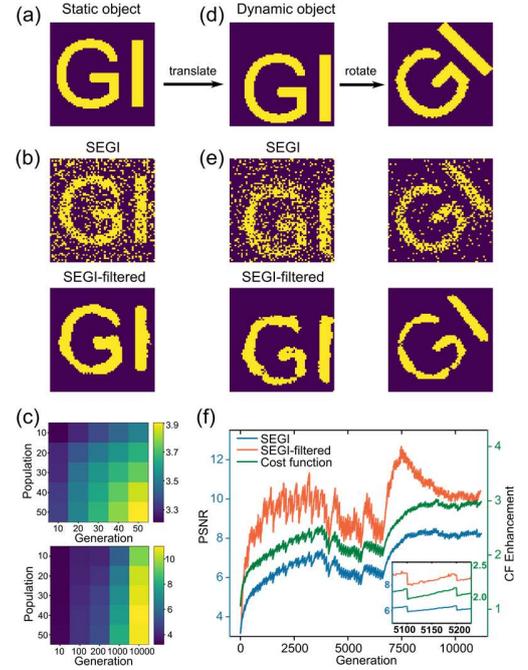

Fig. 2. Numerical results of $k = 1$ SEGI for a binary object. (a) Original object. (b) SEGI raw (top) and median filtered results (bottom) after $G = 1000$ in a population of 30. (c) PSNR of SEGI raw results with (top) a population number range from 10 to 50 and (bottom) a generation number range from 10 to 10000. (d) Example of object frames (52nd and 112th) in a 112-frame image series and corresponding SEGI raw and median filtered results (e) with $G = 100$ and 30 population patterns for each frame, followed by the growths of entire PSNR and cost function values. (f) The PSNR and CF enhancement values of the dynamic SEGI.

Fig. 2 overviews our simulations of SEGI for static and dynamic binary objects. Fig. 2(b) shows the raw and filtered SEGI image of a static binary object (Fig. 2(a)) with N = 30 and G = 1000. By applying a median filter with a 3 × 3 kernel to the retrieved image, we can eliminate the salt-and-pepper noise, preserve the object's edges and improve its peak signal-to-noise ratio (PSNR) from 6.95dB to 14.08dB. To explore the contribution of $N$ and $G$ in SEGI, we plot the PSNR of the raw SEGI images for increasing $N$ and $G$ in Fig. 2(c). The result shows higher values in the upper-right part, which indicates $G$ has higher impact than $N$ with the same $N \times G$ number in the evolution process.

Fig. 2(e) shows the SEGI image of an object that is dynamically changing through translation and rotation (Fig. 2(d)). The movement has been split to 112 individual frames to be imaged. Strikingly, with a limited sampling number of $G = 100$ and $N = 30$, SEGI can still create a recognizable image of the translated object in frame 52; the image quality (PSNR 6.37 for the unfiltered image) is comparable with that for imaging static object under $G = 1000$ and $N = 30$. This enhancement comes from the continuity of object structures along the time dimension, as the output image from the previous frame can serve as a priori knowledge to guide image evolution in the next frame. As a result, the image quality increaseses with the tracking time/frame, and the required sampling ratio is greatly reduced. The PSNR reaches as high as 8.24dB (raw) at frame 112 with a sampling ratio of ~37%.

Fig. 2(f) shows the PSNR enhancement during the entire movement. Under small translations (frames 1-40), the latter frame will inherit image information from the former, which allows SEGI to enhance the PSNR from 3.16dB to about 7dB. The CF value, correlating to the image quality, shows the same trend with PSNR (raw), rising from 1 to 3. When the displacement is larger (frames 41-68), the newly generated images for each frame inherited less information from previous frames. In this case, the PSNR shows a slightly decreasing trend, even though the PSNR increased during the 100 generations for each frame (inset, Fig. 2(f)). With small rotation displacement (frames 69-112), the PSNR increases with the same trend as in small translations. Note that the filtering process can substantially enhance the PSNR for the rotation object.

that in Fig 2. (d) PSNR of SEGI results for static and dynamic imaging in the evolution processes. The population number is 30 in (a)-(d).

We further investigate the role of the weight coefficient $k$ in the image quality. Fig. 3(a) shows the SEGI images of a static object with $k = 2, 3,$ and 4, where increasing $k$ results in a faster grain-filling rate shown than that in Fig. 2(b) with same $N$ and $G$, while the noisy pixels also increase significantly. This phenomenon is more obvious for higher generations (see Supplement 1, Fig. S1). The reason is that an increased $k$ increasingly highlights the contribution of '1' pixels located inside the object. The SEGI images of a dynamic object (Fig. 3(b)) shows the same phenomenon, and the filtered image (Fig. 3(c)) indicates the image dilation tendency appears to decrease image quality. Fig. 3(d) compares the PSNR for $k = 1, 2, 3$ and 4, for both static and dynamic object (see Supplement 1, Fig. S2 for the filtered results). When the generations are less than 3800, the $k = 1$ cost function yields better image quality due to its faster growth rate than the other three for static images, and its PSNR value is generally higher than that for $k = 2, 3, 4$. When the generation number is beyond 3800, $k = 2$ becames the best option for the static object. In terms of dynamic imaging, the superiority of $k = 1$ under small generation numbers gives much higher image quality.

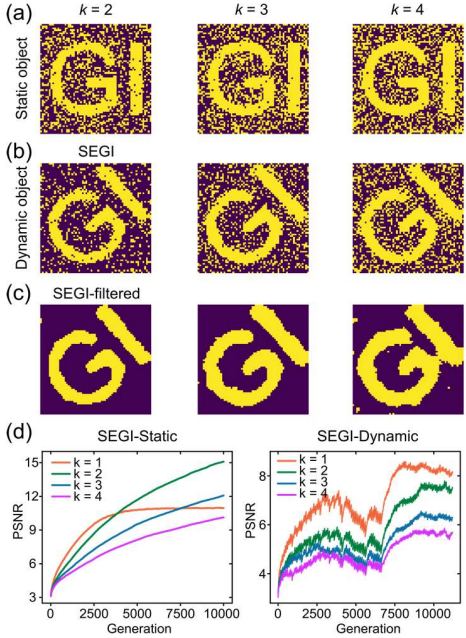

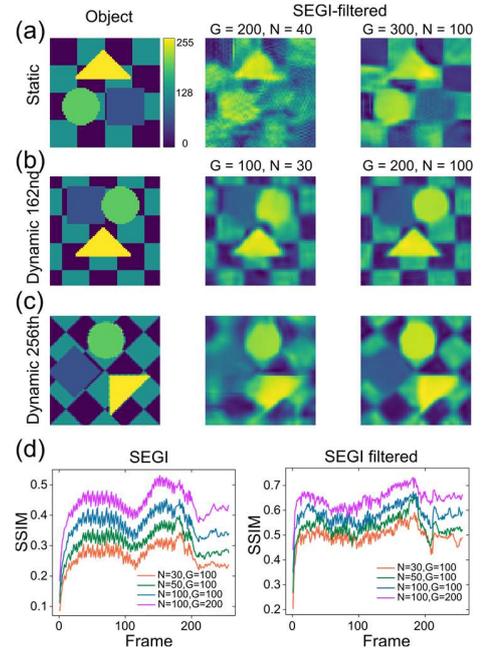

Fig. 3. Characteristic of different $k$ values in the designed cost function. (a) SEGI raw results for static imaging after 1000 generations when $k = 2, 3, 4$. (b) SEGI raw results of 112th frame for dynamic imaging with different $k$ value with 100 generations for each frame and corresponding median filtered images (c). The movement is the same as

Fig. 4. Numerical results of SEGI for a grayscale object. (a) Original object image with coded pseudocolor (left) and filtered SEGI results with $G = 200, N = 40$ (middle) and $G = 300$ and $N = 100$ (right). (b) and (c) show the 162nd and 256th frames selected in a 256-frame object image series (left) and corresponding filtered SEGI results with $G = 100, N = 30$ (middle) and $G = 200$ and $N = 100$ (right). (d) SSIM values of raw and filtered SEGI results for dynamic imaging.

We further employ SEGI to capture 64x64- pixel, 8-bit grayscale images, shown in Fig. 4. In order to balance the numerator and denominator and involve comparable contribution of detected light intensities and pattern weights, we modify the cost function for grayscale imaging as

$$CF(i,g) = \frac{\left(S_i^g\right)^2 \cdot \langle \sum I_i^1(x,y) \cdot I_i^1(x,y) \rangle}{\langle \left(S_i^1\right)^2 \rangle \cdot \sum I_i^g(x,y) \cdot I_i^g(x,y)} \quad (4)$$

Here we choose $k = 2$, as other $k$ values overweight contributions of either large or small pixel values, leading to binary-like image results. Fig. 4(a) shows our grayscale object made of three blocks with different shapes above a chequered background; the color indicates grayscale value. We use the BM3D algorithm for noise reduction [23] and the structural similarity index measure (SSIM, higher is better) to quantitatively analyze the SEGI images. SSIM can perceive changes in structural information between the objects, which is more sensitive to human perception than the change of pixels' absolute values. Two filtered SEGI results with $G = 200, N = 40$ and $G = 300$ and $N = 100$, are shown in Fig. 4(a) middle and right, with sampling ratios of 99% and 369%, respectively. The standard deviation of the filter noise is set to 60 in the filter. A detailed study of the SEGI raw and filtered images is shown in Supplement 1 Section 2, Fig. S4 & Fig. S5.

For dynamic imaging, we extend the object image to a 256-frame image series by translating and rotating the elements and background (see Supplement 1, Fig. S6 for the details). The 162nd and 256th original images are shown in the left of Fig. 4(b) and Fig. 4(c). The middle images are generated under $G = 100$ with $N = 30$. The right images are generated under $G = 200$ with $N = 100$. Thus, the sampling ratios are 36.6% and 244.1%, respectively. The rough structures in the object image are retrieved in the middle SEGI images with the sampling ratios of 36.6%, while for the static case it needs a sampling ratio of about 368.7%. The SSIM values of dynamic SEGI imaging under $G = 100, N = 30$; $G = 100, N = 50$; $G = 100, N = 100$ and $G = 200, N = 100$ are shown in Fig. 4(d). The image quality could be further improved by increasing $G$ or $N$ but at a cost of a longer sampling time. More details are provided in Supplement 1 Fig. S7 to Fig. S12.

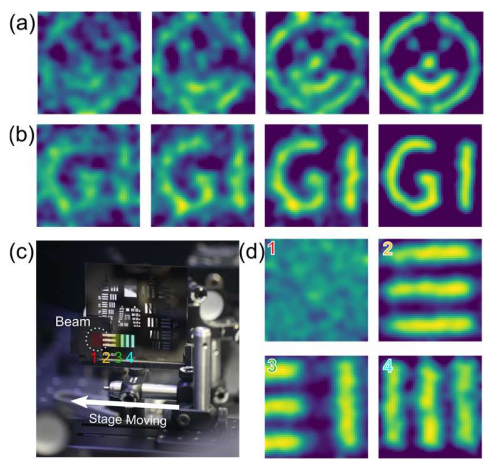

Fig. 5. Experimental results of SEGI. SEGI for static objects (~10mm masks) with $N = 20$ and $G = 100, 200, 500, 2000$ are shown in (a) and (b) from left to right, respectively, and $k = 2$. For dynamic SEGI, a 1951 USAF test target mounted on a translation stage (c) is used to move across the FOV. Four selected frames of dynamic SEGI results are shown in (d), which corresponding to the indicated positions in (c).

To verify this concept, we use a DMD (ViALUX GmbH V-7001) for pattern projection, collecting light from the object on a silicon photodiode. Custom python-based algorithms are used to generate and update DMD patterns. To speed up data transfer to the DMD, we use a 48x48- pixel area on the DMD for SEGI, with an illumination time of 94μs/frame and 25ms per generation. Two 3D printed transparent masks (~10mm 'smiling face' and letters 'GI') act as the target. SEGI results with a Gaussian blur filter are shown in Fig. 5(a) and Fig. 5(b) with $G$ = 100, 200, 500, 2000, respectively, and $N$ = 20. The object outline appears when $G = 500$, corresponding to a sampling ratio of 218%.

We also validate SEGI by imaging a moving object (a 1951 USAF target (R3L3S1N, Thorlabs)) mounted on a linear translation stage (DDSM100, Thorlabs) shown in Fig. 5(c)) with a translation speed of 50μm/s. Four selected frames of dynamic SEGI when the target is moving are shown in Fig. 5(d). Details are provided in Supplement 1 Section 3.

In conclusion, we present a new class of ghost imaging, SEGI, that allows image formation of objects in the display patterns of the spatial light modulator without post-reconstruction. A genetic algorithm is used to generate 'self-evolving patterns'. Other optimization algorithms or machine learning based networks could be implemented to further enhance SEGI. Our technique can physically project the object image in real time by beamspliting the outgoing light from the illuminating spatial light modulator, which could be used for adaptive/smart illumination for bioimaging or studio lighting. We anticipate that this work will open opportunities for developing integrated single-pixel cameras and novel ghost imaging based modalities.

**Funding.** UTS Faculty of Engineering and IT. The Australian Research Council DECRA fellowship (DE200100074, F.W.) and China Scholarship Council (B.L. : No.201706020170; C.C. : No. 201607950009).

See Supplement 1 for supporting content.